\documentstyle[12pt]{article}

\begin{document}
\setcounter{page}{1}
\pagestyle{plain} \vspace{1cm}
\begin{center}
\Large{\bf Comparison of Frames: Jordan vs Einstein Frame for a Non-minimal Dark Energy Model}\\
\small \vspace{1cm}
{\bf Kourosh Nozari$^{a,b}$}\quad\ and \quad {\bf S. Davood Sadatian$^{c}$}\\
\vspace{0.5cm} {\it $^{a}$Department of Physics,
Faculty of Basic Sciences,\\
University of Mazandaran,\\
P. O. Box 47416-95447,
Babolsar, IRAN\\
$^{b}$Research Institute for Astronomy and Astrophysics of Maragha,
\\P.
O. Box 55134-441, Maragha, IRAN \\
and\\
$^{c}$Islamic Azad University, Nishabour Branch, Nishabour, IRAN}\\

{\it knozari@umz.ac.ir}\\
{\it d.sadatian@umz.ac.ir}
\end{center}
\vspace{1.5cm}
\begin{abstract}
We construct a dark energy model where a scalar field non-minimally
coupled to gravity plays the role of the dark component. We compare
cosmological consequences of this non-minimal coupling of the scalar
field and gravity in the spirit of the dark energy paradigm in
Jordan and Einstein frames. Some important issues such as phantom
divide line crossing, existence of the bouncing solutions and the
stability of the solutions are compared in these two frames. We show
that while a non-minimally coupled scalar field in the Jordan frame
is a suitable dark energy component with capability to realize
phantom divide line crossing, its conformal transformation in the
Einstein frame has not this capability. The conformal transformation
from Jordan frame to Einstein frame transforms the equation of state
parameter of the dark energy component to its minimal form with a
redefined scalar field and in this case it is impossible to realize
a phantom phase with possible crossing of the phantom divide line.\\
{\bf PACS}: 98.80.-k, 98.80.Cq\\
{\bf Key Words}: Scaler-Tensor Theories, Dark Energy Models,
Bouncing Solutions
\end{abstract}
\vspace{1.5cm}
\newpage

\section{Introduction}
Recently, evidences from supernova searches data [1,2], cosmic
microwave background (CMB) results [3-5] and also Wilkinson
Microwave Anisotropy Probe (WMAP) data [6,7], indicate an positively
accelerating phase of the cosmological expansion today and this
feature shows that the simple picture of universe consisting of a
pressureless fluid is not enough. In this regard, the universe may
contain some sort of additional negative-pressure dark energy.
Analysis of the five year WMAP data [8-11] shows that there is no
indication for any significant deviations from Gaussianity and
adiabaticity of the CMB power spectrum and therefore suggests that
the universe is spatially flat to within the limits of the
observational accuracy. Further, the combined analysis of the
three-year WMAP data with the supernova Legacy survey (SNLS) [8],
constrains the equation of state $w_{de}$, corresponding to almost
${74\%}$ contribution of dark energy in the currently accelerating
universe, to be very close to that of the cosmological constant
value. Moreover, observations appear to favor a dark energy equation
of state, $w_{de}<-1$ [12,13]. Therefore a viable cosmological model
should admit a dynamical equation of state that might have crossed
the value $w_{de}= -1$, in the recent epoch of cosmological
evolution. In fact, to explain positively accelerated expansion of
the universe, there are two alternative approaches: incorporating an
additional cosmological component to the matter part of the Einstein
equations or modifying gravity ( the geometric part of the Einstein
field equations) at cosmological scale. Multi-component dark energy
with at least one non-canonical phantom field is a possible
candidate of the first alternative. This viewpoint has been studied
extensively in literature ( see [14] and references therein ).

On the other hand, despite of several successes, the standard model
of cosmology suffers from a series of problems. The most serious of
these problems is the problem of initial singularity because the
laws of physics break down at the singularity point. In order to
avoid this lawlessness, there is a huge interest in the solutions
that do not display divergencies. These solutions could be obtained
at a classical level or by quantum modifications. Most of the
efforts in quantum gravity is devoted to reveal the nature of the
initial singularity and to understand the origin of matter,
non-gravitational fields, and the very nature of the spacetime. In
recent analysis done within the loop quantum cosmology, the Big Bang
singularity is replaced by a quantum \emph{Big Bounce} with finite
energy density of matter. This scenario has strong quantum effects
at the Planck scale too. Another motivation to remove the initial
singularity is the initial value problem. A sound gravitational
theory needs to have a well-posed Cauchy problem. Due to the fact
that the gravitational field diverges at the singularity, we could
not have a well-formulated Cauchy problem as we cannot set the
initial values at that point.

With these preliminaries, the purpose of the present paper is to
study some currently important cosmological issues such as phantom
divide line crossing, avoiding singularities by realization of the
bouncing solutions and the stability of these solutions in a
non-minimally coupled scalar field model of universe in Jordan and
Einstein frames and in the spirit of dark energy model. We compare
cosmological consequences of the non-minimal coupling between scalar
field and gravity in the spirit of the dark energy scenario in these
two frames. Some important issues such as phantom divide line
crossing, existence of the bouncing solutions and the stability of
the solutions are compared in these two frames. Especially, we
analyze the parameter space of the model numerically to show that
which frame allows for stability of the solutions in the separate
regions of the $\omega-\omega^{'}$ phase-plane. We show that with
only one scalar field non-minimally coupled to gravity, crossing of
the phantom divide line can be realized just in the Jordan frame. By
transforming to Einstein's frame, we show that this model cannot
account for crossing of the phantom divide line. We show that while
a non-minimally coupled scalar field in the Jordan frame is a
suitable dark energy component with capability to realize phantom
divide line crossing, its conformal transformation in the Einstein
frame has not this capability. In fact, conformal transformation
from Jordan frame to Einstein frame transforms the equation of state
parameter of dark energy component to its minimal form with a
redefined scalar field and in this case it is impossible to realize
a phantom phase with possible crossing of the phantom divide line.

\section{A Ricci-Coupled Scalar Field Model in the Jordan Frame }
For a model universe with a non-minimally coupled scalar field as
matter content of the universe, the action in the absence of other
matter sources in the Jordan frame can be written as follows
\begin{equation}
S=\int d^{4}x\sqrt{-g}\bigg[\frac{1}{k_{4}^{2}}\alpha(\phi)
R[g]-\frac{1}{2} g^{\mu\nu} \nabla_{\mu}\phi\nabla_{\nu}\phi
-V(\phi) \bigg],
\end{equation}
where we have included an explicit and general non-minimal coupling
of the scalar field and gravity as $\alpha(\phi)$. For simplicity,
from now on we set ${k_{4}}^{2}\equiv8\pi G_{N}=1$. Variation of the
action with respect to the metric gives the Einstein equations
\begin{equation}
R_{\mu\nu}-\frac{1}{2}g_{\mu\nu}R=\alpha^{-1}{\cal{T}}_{\mu\nu}.
\end{equation}
${\cal{T}}_{\mu\nu}$, the energy-momentum tensor of the scalar field
non-minimally coupled to gravity, is given by
\begin{equation}
{\cal{T}}_{\mu\nu}=\nabla_{\mu}\phi\nabla_{\nu}\phi-\frac{1}{2}g_{\mu\nu}(\nabla\phi)^{2}-g_{\mu\nu}V(\phi)+
g_{\mu\nu}\Box\alpha(\phi)-\nabla_{\mu}\nabla_{\nu}\alpha(\phi),
\end{equation}
where $\Box$ shows 4-dimensional d'Alembertian. For FRW universe
with line element defined as
\begin{equation}
ds^{2}=-dt^{2}+a^{2}(t)d{\Sigma_{k}}^{2},
\end{equation}
where $d{\Sigma_{k}}^{2}$ is the line element for a manifold of
constant curvature $k = +1,0,-1$, the equation of motion for scalar
field $\phi$ is
\begin{equation}
\nabla^{\mu}\nabla_{\mu}\phi=V'-\alpha'R[g],
\end{equation}
where a prime denotes derivative of any quantity with respect to\,
$\phi$. This equation can be rewritten as
\begin{equation}
\ddot{\phi}+3\frac{\dot{a}}{a}\dot{\phi}+\frac{dV}{d\phi}=
\alpha'R[g].
\end{equation}
where a dot denotes the derivative with respect to cosmic time\, $t$
\, and Ricci scalar is given by
\begin{equation}
R=6\bigg(\dot{H}+2H^{2}+\frac{k}{a^{2}}\bigg).
\end{equation}
With this non-minimally coupled scalar field as matter content of
the universe, cosmological dynamics are described by
\begin{equation}
\frac{\dot{a}^{2}}{a^{2}}=-\frac{k}{a^{2}}+\frac{\rho}{3},
\end{equation}
and
\begin{equation}
\frac{\ddot{a}}{a}=-\frac{1}{6}(\rho+3p).
\end{equation}
In these equations, the effect of the non-minimal coupling of the
scalar field and gravity is hidden in the definition of $\rho$ and
$p$. We assume that scalar field, $\phi$, has only time dependence
and using (3), we find
\begin{equation}
\rho=\alpha^{-1}\bigg(\frac{1}{2}\dot{\phi}^{2}+V(\phi)-6\alpha'H\dot{\phi}\bigg),
\end{equation}
\begin{equation}
p=\alpha^{-1}\bigg(\frac{1}{2}\dot{\phi}^{2}-V(\phi)+
2\Big(\alpha'\ddot{\phi}+2H\alpha'\dot{\phi}+\alpha''\dot{\phi}^2\Big)\bigg),
\end{equation}
where $H=\frac{\dot{a}}{a}$ is Hubble parameter. Now, equation (9)
takes the following form
\begin{equation}
\frac{\ddot{a}}{a}=-\frac{1}{6}\alpha^{-1}\bigg(2\dot{\phi}^{2}-2V(\phi)+
6\Big(\alpha'\ddot{\phi}+H\alpha'\dot{\phi}+\alpha''\dot{\phi}^2\Big)\bigg),
\end{equation}
and dynamics of the equation of state parameter is given by
\begin{equation}
w\equiv\frac{p}{\rho}=\frac{\dot{\phi}^{2}-2V(\phi)+
4\Big(\alpha'\ddot{\phi}+2H\alpha'\dot{\phi}+\alpha''\dot{\phi}^2\Big)}
{\dot{\phi}^{2}+2V(\phi)-12\alpha'H\dot{\phi}}.
\end{equation}
From this equation, when $\dot{\phi}=0$, we obtain $p=-\rho$. In
this case $\rho$ is independent of $a$ and $V(\phi)$ plays the role
of a cosmological constant. In the minimal case when
$\dot{\phi}^{2}< V(\phi)$, using (9) we obtain $p<-\frac{\rho}{3}$
which shows an accelerated expansion which is driven by cosmological
constant. However, cosmological constant is not a good candidate for
dark energy since its suffers from several conceptual problems such
as its unknown origin and also need to huge amount of fine-tuning.
In non-minimal case the cosmological dynamics depends on the value
of the non-minimal coupling. In order to solve the Friedmann
equation (8), we choose the following scalar field potential
\begin{equation}
V(\phi)=\frac{1}{2}m^2\phi^2
\end{equation}
With this potential, a possible solution of our basic equations,
(6), (8) and (10) is as follows ( see [15] for a similar argument)
\begin{equation}
\phi=\sqrt{C_0}cos(mt)
\end{equation}
where $C_0$ is a parameter with the dimension of mass squared
describing the oscillating amplitude of the fields. Here we assume
$\alpha(\phi)=\frac{1}{2}(1-\xi\phi^2)$. We note that there are two
critical values of $\phi$ given by
$\phi_{c}=\pm\frac{1}{\sqrt{\xi}}$ that should be avoided to have
well-defined field equations. Using (14) and (15) in (10), we find
\begin{equation}
\rho=\frac{m^2C_{0}-12\xi HC_0msin(mt)cos(mt)}{1-\xi
C_{0}cos^2(mt)},
\end{equation}
where those values of the cosmic time coordinates that lead to
singular energy density are excluded from our considerations.
Therefore, for a flat spatial geometry, Friedmann equation (8) can
be rewritten as follows
\begin{equation}
H={\frac{\left(6\cos\left(mt\right)\sin\left(mt\right)C_0\xi\pm\sqrt
{36\left(\cos\left(mt\right)\right)^{2}\left(\sin
 \left(mt\right)\right)^{2}{C_0}^{2}{\xi}^{2}+C_0-\xi{C_0}^{2}
 \left(\cos\left(mt\right)\right)^{2}}\right)m}{-1+\xi C_0
\cos^{2}\left(mt\right)}}
\end{equation}
We avoid imaginary values of the Hubble parameter in which follows.

\subsection{Crossing of the phantom divide line}
Non-minimal coupling of the scalar field and gravity in the Jordan
frame provides a suitable framework for explanation of the late-time
accelerated expansion [16]. On the other hand, after substituting
corresponding relations for $\phi$, $H$ and $V$ in equation (13),
dynamics of the equation of state parameter for a non-minimally
coupled scalar field is given by
$$\omega(t)=\Bigg\{-8\xi\epsilon \sin mt\cos mt
\sqrt{-36\left(-\frac{1}{36}+C_0{\xi}^{2}\cos^{4} mt-\xi
C_0\left(\xi-\frac{1}{36}\right)\cos^{2}mt\right)C_0}$$
$$+48\xi C_0\left(-\frac{1}{24}+\xi\right)\cos^{4}mt+
\left(2+\xi C_0-52C_0{\xi}^{2}\right)\cos^{2}mt- 1+4\xi\Bigg\}\times$$
$$\Bigg[-12\xi\epsilon\sin mt\cos mt \sqrt{-36\left(-\frac{1}{36}+C_0 {\xi}^{2}\cos^{4} mt
-\xi C_0\left(\xi-\frac{1}{36}\right) \cos^{2}mt \right)C_0}-1$$
\begin{equation}
+72C_0 {\xi}^{2}\cos^{4}mt -72\xi\left(-{\frac{1}{72}}+\xi\right)
C_0\cos^{2}mt\Bigg]^{-1}
\end{equation}
where $\epsilon=\pm 1$. Figure $1$ shows the crossing of the phantom
divide line with equation of state parameter of this non-minimally
coupled scalar field. Nevertheless, as figure $2$ shows, in the case
of $\xi=0$, that is a single and minimally coupled scalar field,
there is no crossing of the phantom divide line, as has been
emphasized by other literature such as [14]. We note that our
analysis shows that for negative values of $\xi$ ( corresponding to
anti-gravitation) and any values of $\epsilon$ ( $\epsilon=-1$ or
$\epsilon=+1$), it is impossible to realize the phantom divide line
crossing in this setup.

\begin{figure}
\begin{center}\includegraphics{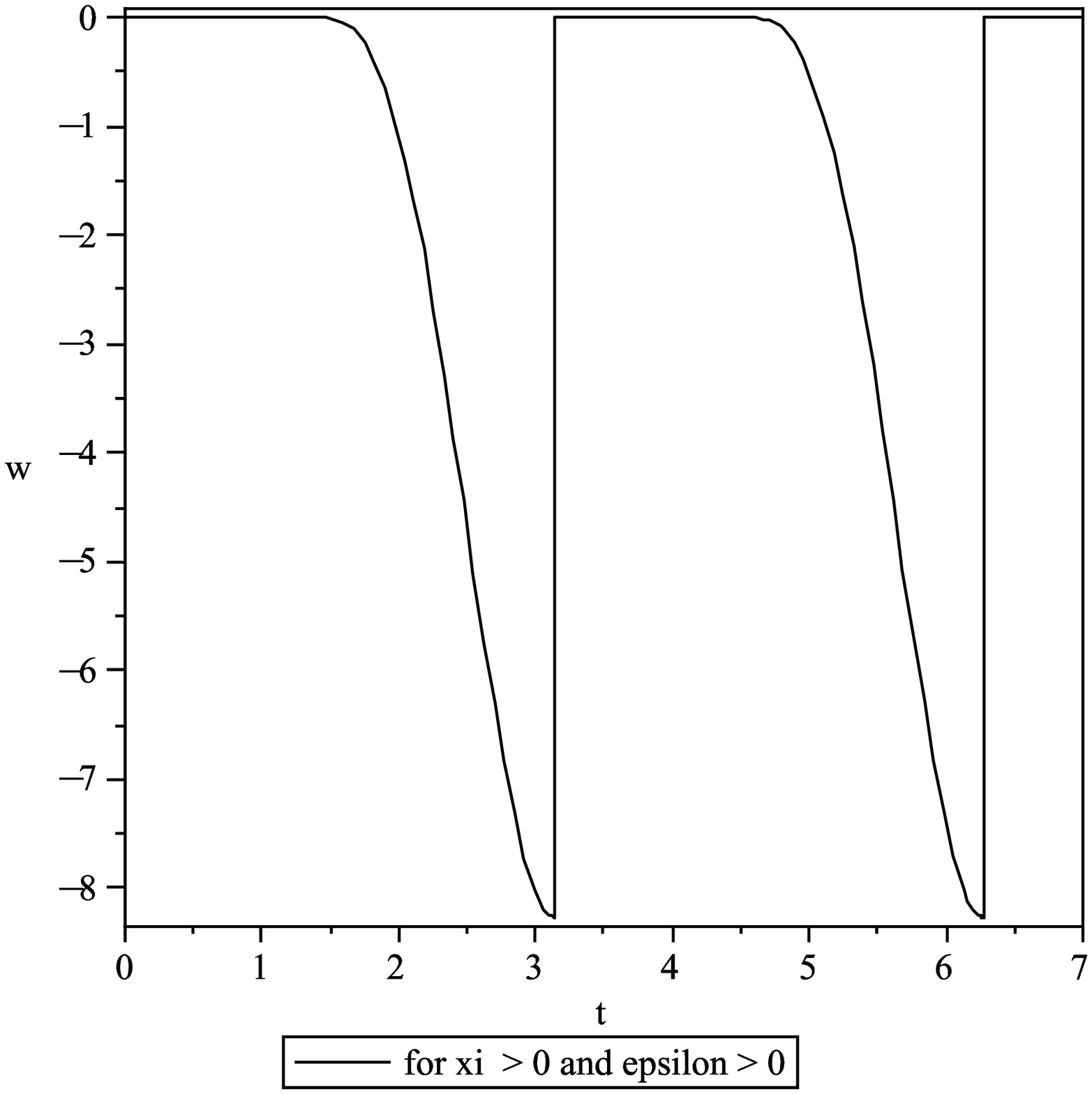} \vspace{5.5cm}\includegraphics{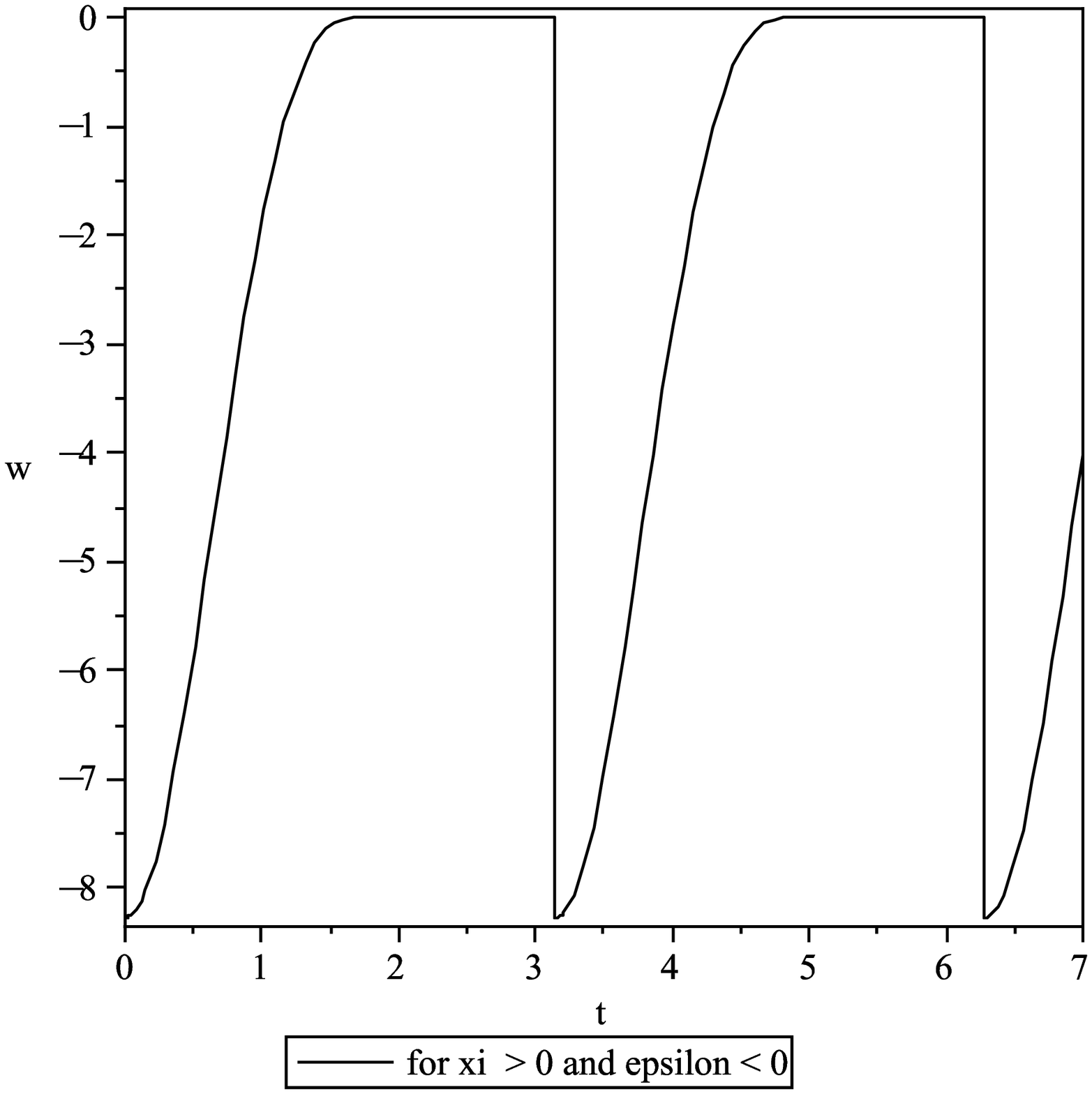}
\end{center}
\caption{\small {A non-minimally coupled scalar field in Jordan
frame has the capability to realize crossing of the phantom divide
line by its equation of state parameter in a suitable subspace of
the model parameter space.}}
\end{figure}

\begin{figure}[htp]
\begin{center}\includegraphics{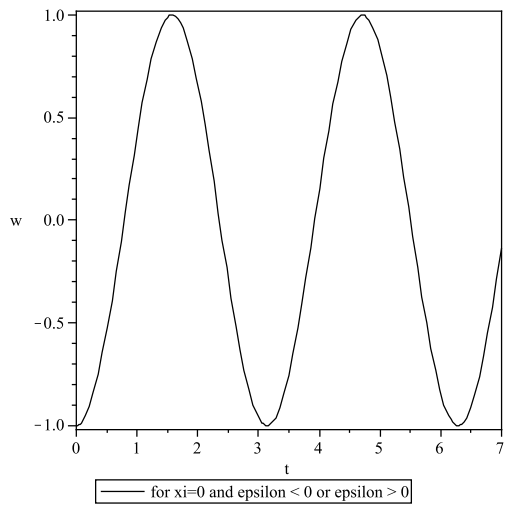} \vspace{5.3cm}
\end{center}
 \caption{\small {Equation of state parameter of a single, minimally coupled scalar
 field ( with $\xi=0$ ), cannot explain crossing of the phantom divide line.}}
\end{figure}

\subsection{Bouncing behavior of the model}
A bouncing universe with an initial contraction to a non-vanishing
minimal radius and a subsequent expanding phase provides a possible
solution to the singularity problem of the standard Big Bang
cosmology. For a successful bounce, it can be shown that within the
framework of the standard 4-dimensional Friedmann-Robertson-Walker
(FRW) cosmology with Einstein gravity, the null energy condition
(NEC) is violated for a period of time around the bouncing point.
Moreover, for the universe entering into the hot Big Bang era after
the bouncing, the EoS of the matter content $w$ in the universe must
transit from $w<-1$ to $w>-1$ [17].

The model proposed to understand the behavior of dark energy with an
EoS of $w>-1$ in the past and $w<-1$ at present, has been supported
by the observational data [18]. This is a dynamical model of dark
energy and it differs from the cosmological constant, Quintessence,
Phantom, K-essence and so on in the determination of the
cosmological evolution. Here we study the possibility of realization
of the bouncing solution in a model universe dominated by the
nonminimally coupled scaler field as matter content. We start with a
detailed examination of the necessary conditions required for a
successful bounce. During the contracting phase, the scale factor
$a(t)$ is decreasing, i.e., $\dot a(t)<0$, and in the expanding
phase we have $\dot a(t)>0$. At the bouncing point, $\dot a(t)=0$,
and around this point $\ddot a(t)>0$ for a period of time.
Equivalently, in the bouncing cosmology the Hubble parameter $H$
runs across zero from $H<0$ to $H>0$ and $H=0$ at the bouncing
point. In our model this behavior is shown in figure $3$. A
successful bounce requires that around this point
\begin{equation}
\dot H=-4\pi G\rho (1+w)> 0~.
\end{equation}
From equation (18) one can see that $w<-1$ in a neighborhood of the
bouncing point. After the bounce the universe needs to enter into
the hot Big Bang era, otherwise the universe filled with the matter
with an EoS $w<-1$ will reach the big rip singularity as what
happens to the Phantom dark energy which violates the null energy
condition [19]. This requires the EoS of the matter to transit from
$w<-1$ to $w>-1$.

By solving the Friedmann equation (17), we can study the scale
factor versus the cosmic time, $t$. If we integrate equation (17),
we find
\begin{equation}
\tiny
 a(t)=a_0exp\Bigg[ \int
{\frac{\left(6\cos\left(mt\right)\sin\left(mt\right)C_0\xi\pm\sqrt
{36\left(\cos\left(mt\right)\right)^{2}\left(\sin
 \left(mt\right)\right)^{2}{C_0}^{2}{\xi}^{2}+C_0-\xi{C_0}^{2}
 \left(\cos\left(mt\right)\right)^{2}}\right)m}{-1+\xi C_0
 \left(\cos\left(mt\right)\right)^{2}}}dt\Bigg]
\end{equation}

\begin{figure}
\begin{center}\includegraphics{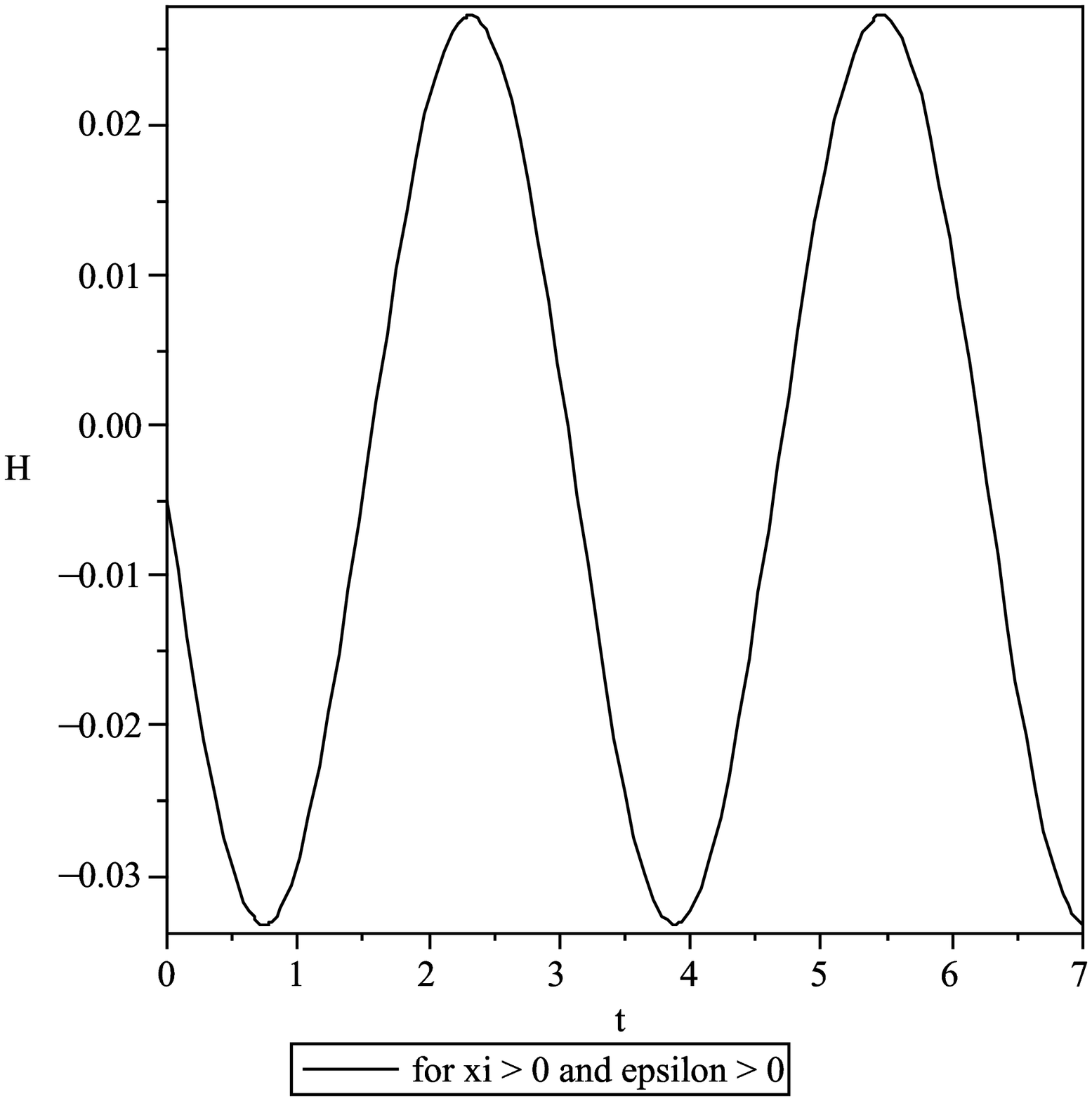} \vspace{5.5cm}\includegraphics{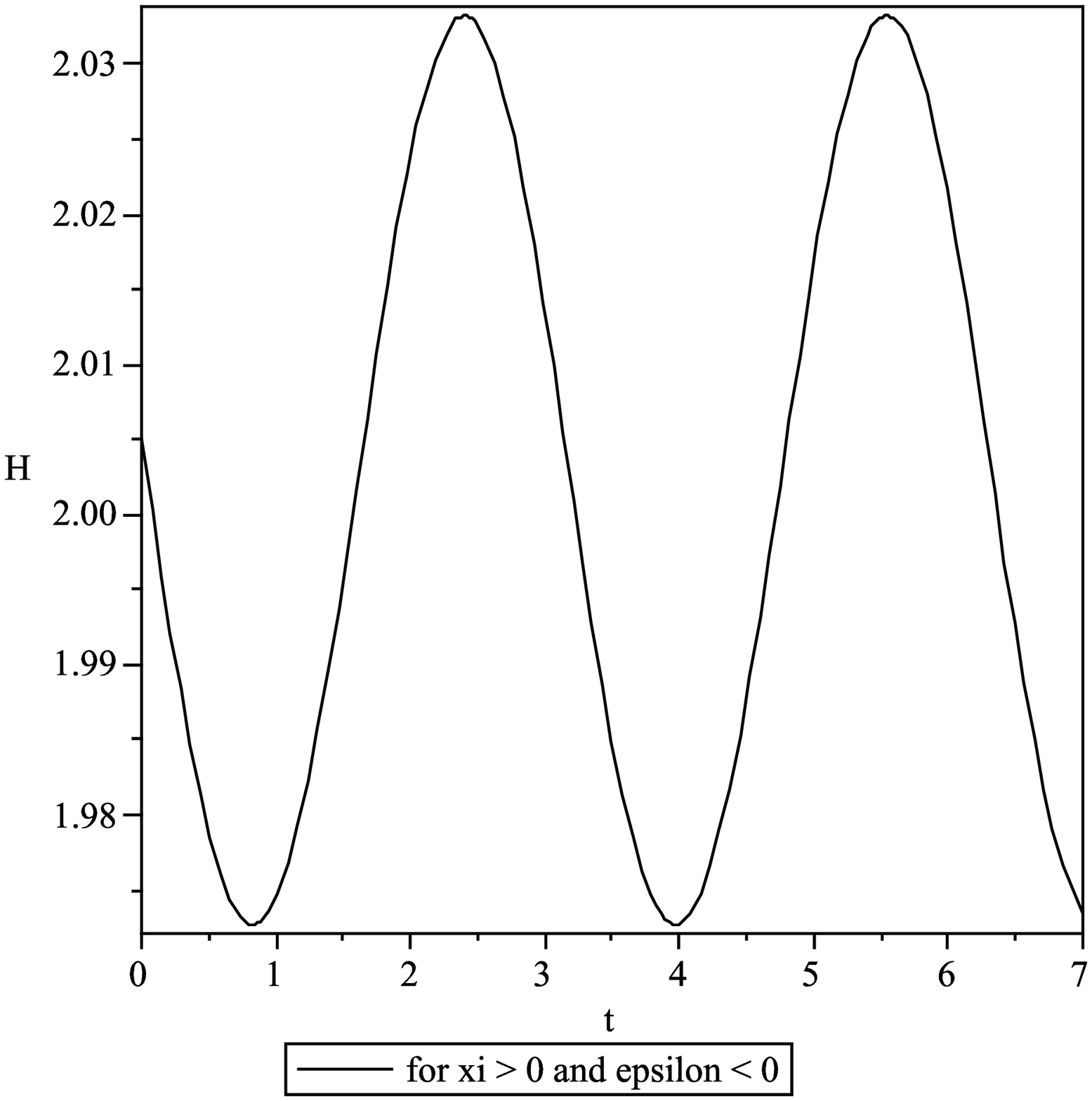}
\end{center}
\caption{\small {Variation of the Hubble parameter $H$ relative to
cosmic time $t$. The left hand side figure shows that the universe
can switches alternatively between expanding and contracting
phases.}}
\end{figure}
Using equation (20), we plot the behavior of the scale factor versus
the cosmic time.
\begin{figure}[htp]
\begin{center}\includegraphics{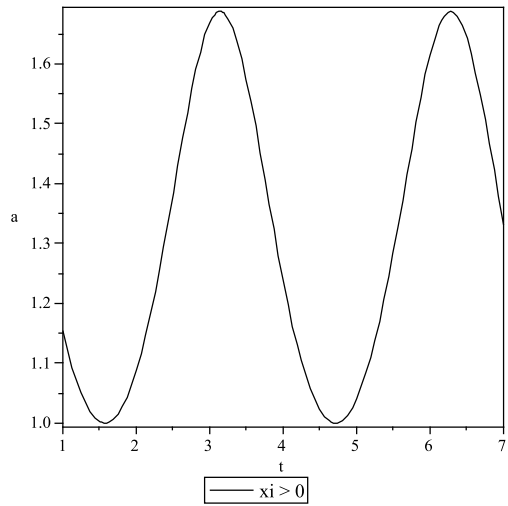} \vspace{5.3cm}
\end{center}
 \caption{\small {Variation of the scale factor $a$ relative to cosmic
time $t$. Note that the initial conditions have been set so that
integrand of equation (20) has non-vanishing derivative at the
minimum point.}}
\end{figure}
One can see from Figures $3$ (left) and $4$ that a nonsingular
bounce happens when the Hubble parameter $H$ runs across zero with a
minimal and non-vanishing scale factor $a$.

\subsection{Stability of the model}
Now we study the stability of our model. In order to study the
classical stability of our model, we analyze the behavior of the
model in the $\omega-\omega' $ plane where $\omega'$ is the
derivative of $\omega$ with respect to the logarithm of the scale
factor ( see [20-23] for a similar analysis for other interesting
cases)
\begin{equation}
\omega'\equiv\frac{d\omega}{d\ln a}=\frac{d\omega}{dt}\frac{dt}{d
\ln a}=\frac{\dot{\omega}}{H}.
\end{equation}
The sound speed expresses the phase velocity of the inhomogeneous
perturbations of the quintessence field. We define the function
$c_a$ as
\begin{equation}
{c_a}^2\equiv\frac{\dot{p}}{\dot{\rho}}
\end{equation}
If the matter is considered as a perfect fluid, this function would
be the adiabatic sound speed of this fluid. We note that with scalar
fields that do not obey perfect fluid form necessarily, this
quantity is not actually a sound speed. In which follows, we demand
that ${c_a}^2> 0$ in order to avoid the big rip singularity at the
end of the universe evolution. The conservation of the quintessence
field effective energy density can be stated as
\begin{equation}
\frac{d\rho_{quintessence}}{dt}+3H(\rho_{quintessence}+p_{quintessence})=0
\end{equation}
Since the dust matter obeys the continuity equation and the Bianchi
identity keeps valid, total energy density satisfies the continuity
equation. From above equation, we have
\begin{equation}
\dot{\rho}_{de}=-3H\rho_{de}(1+\omega_{de})
\end{equation}
Using equation of state $p_{de}=\omega_{de}\rho_{de}$, we obtain
\begin{equation}
\dot{p}_{de}=\dot{\omega}\rho_{de}+\omega_{de}\dot{\rho}_{de}
\end{equation}
Therefore, the function ${c_a}^2$ could be rewritten as
\begin{equation}
{c_a}^2=\frac{\dot{\omega}_{de}}{-3H(1+\omega_{de})}+\omega_{de}
\end{equation}
In this situation, the $\omega-\omega'$ plane is divided into four
regions defined as follows
$$I:~~~\omega_{de} >
-1,~~~~~\omega'>3\omega(1+\omega)~~\Rightarrow~~
{c_a}^2 > 0$$
$$II:~~~\omega_{de} >
-1,~~~~~\omega'<3\omega(1+\omega)~~\Rightarrow~~ {c_a}^2 < 0$$
$$III:~~~\omega_{de} <
-1,~~~~~\omega'>3\omega(1+\omega)~~\Rightarrow~~ {c_a}^2 < 0$$
\begin{equation}
IV:~~~\omega_{de} < -1,~~~~~\omega'<3\omega(1+\omega)~~\Rightarrow~~
{c_a}^2 > 0
\end{equation}
As one can see from these relations, the regions I and IV have the
classical stability in our model. We plot the behavior of the model
in the $\omega-\omega'$  phase plane and identify the regions
mentioned above in figure 5.
\begin{figure}[htp]
\begin{center}\includegraphics{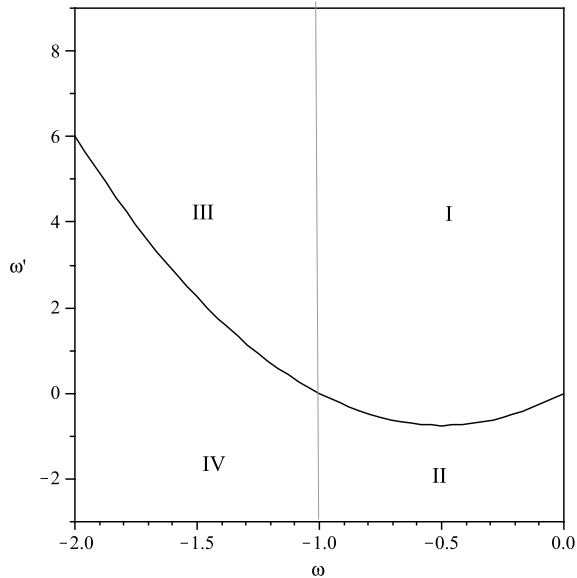} \vspace{8cm}
\end{center}
 \caption{\small {Bounds on $\omega'$  as a function of $\omega$ in $\omega-\omega'$
  phase plane. The stable regions are I and IV.}}
\end{figure}

\section{A Ricci-Coupled Scalar Field in the Einstein Frame }
Now we extend our study of the mentioned cosmological issues to the
Einstein frame by adopting a conformal transformation. The action
(1) in the Jordan frame can be rewritten as follows
\begin{eqnarray}
S=\int d^4x \sqrt{-g} \left[ \frac{1}{2}R
-\frac{1}{2}g^{\mu\nu}\partial_{\mu}\phi \partial_{\nu}\phi
-\frac{1}{2} \xi R \phi^2-V(\phi)\right]
\end{eqnarray}
where we assumed ${k_4}^{2}=1$ and
$\alpha(\phi)=\frac{1}{2}(1-\xi\phi^2)$ and $\xi$ is a non-minimal
coupling. The metric signature convention is chosen to be
$(+\,-\,-\,-)$ with spatially flat Robertson-Walker metric as
follows
\begin{eqnarray}
ds^2=dt^2-a^{2}(t)\delta_{ij}dx^idx^j.
\end{eqnarray}
To obtain the fundamental background equations in Einstein frame, we
perform the following conformal transformation
\begin{eqnarray}
\hat{g}_{\mu\nu}=\Omega g_{\mu\nu}, ~~~~~~ \Omega=1-\xi\phi^2.
\end{eqnarray}
Here we use a hat on a variable defined in the Einstein frame. The
conformal transformation gives ( see for instance [16] and
references therein)
\begin{eqnarray}
S=\int d^4\hat{x} \sqrt{-\hat{g}} \left[ \frac{1}{2}\hat{R}
-\frac{1}{2}F^2(\phi)\hat{g}^{\mu\nu}\partial_{\mu}\phi\partial_{\nu}\phi
-\hat{V}(\phi) \right],
\end{eqnarray}
where by definition
\begin{eqnarray}
F^2(\phi)\equiv\frac{1-\xi\phi^2(1-6\xi)}{(1-\xi\phi^2)^2}
\end{eqnarray}
and
\begin{eqnarray}
\hat{V}(\phi)\equiv\frac{V(\phi)}{(1-\xi\phi^2)^2}.
\end{eqnarray}
Therefore, one may redefine the scalar field as follows
\begin{eqnarray}
\frac{d\hat{\phi}}{d\phi}=F(\phi)=\frac{\sqrt{1-\xi\phi^2(1-6\xi)}}{1-\xi\phi^2}.
\end{eqnarray}
When we investigate the dynamics of universe in the Einstein frame,
we should transform our coordinates system to make the metric in the
Robertson-Walker form
\begin{eqnarray}
\hat{a}=\sqrt{\Omega}a,~~d\hat{t}=\sqrt{\Omega}dt,
\end{eqnarray}
and we obtain
\begin{eqnarray}
d\hat{s}^2=d\hat{t}^2-\hat{a}^2(\hat{t})\delta_{ij}dx^idx^j.
\end{eqnarray}
Note that the physical quantities in Einstein frame should be
defined in this coordinate system. Now the field equations can be
written as follows
\begin{eqnarray}
\hat{H}^2=\frac{1}{3}\left[\frac{1}{2}
\bigg(\frac{d\hat{\phi}}{d\hat{t}}\bigg)^2+\hat{V}(\hat{\phi})
\right]=\frac{\hat{\rho}}{3},
\end{eqnarray}
\begin{eqnarray}
\frac{d^2\hat{\phi}}{d\hat{t}^2}+3\hat{H}\frac{d\hat{\phi}}{d\hat{t}}+
\frac{d\hat{V}}{d\hat{\phi}}=0
\end{eqnarray}
where $\hat{H}=\frac{\hat{\dot{a}}}{\hat{a}}$.

We assume that scalar field $\hat{\phi}$ has only time dependence
and we find dynamics of equation of state as follows
\begin{equation}
\hat{\omega}_{\phi}=\frac{\hat{p}}{\hat{\rho}}=\frac{\frac{1}{2}
\bigg(\frac{d\hat{\phi}}{d\hat{t}}\bigg)^2-\hat{V}(\hat{\phi})}{\frac{1}{2}
\bigg(\frac{d\hat{\phi}}{d\hat{t}}\bigg)^2+\hat{V}(\hat{\phi})}.
\end{equation}
This is an interesting result: while a non-minimally coupled scalar
field in the Jordan frame is a suitable dark energy component with
capability to realize phantom divide line crossing, its conformal
transformation in the Einstein frame has not this capability. In
fact, conformal transformation from Jordan frame to Einstein frame
transforms the equation of state parameter of dark energy component
to its minimal form with a redefined scalar field and in this case
it is impossible to realize a phantom phase with possible crossing
of the phantom divide line. In a minimal coupling case with
$\alpha=0$, one can obtain equation of state parameter
$\omega=1-2cos^2(mt)$. As figure $2$ shows, this minimally coupled
scalar field cannot explain the crossing of the phantom divide line.
However, we note that multi component minimally coupled scalar
fields can realize this crossing [14].

To study bouncing behavior of the solutions in the Einstein frame
with above mentioned redefined scalar field, we see that Hubble
parameter and scale factor have no possibility to run across zero.
This behavior has been shown in figure $6$ using ansatz presented in
the previous section. Therefore, in the Einstein frame we have not
any possibility to realize bouncing solutions with this redefined
scalar field.
\begin{figure}
\begin{center}\includegraphics{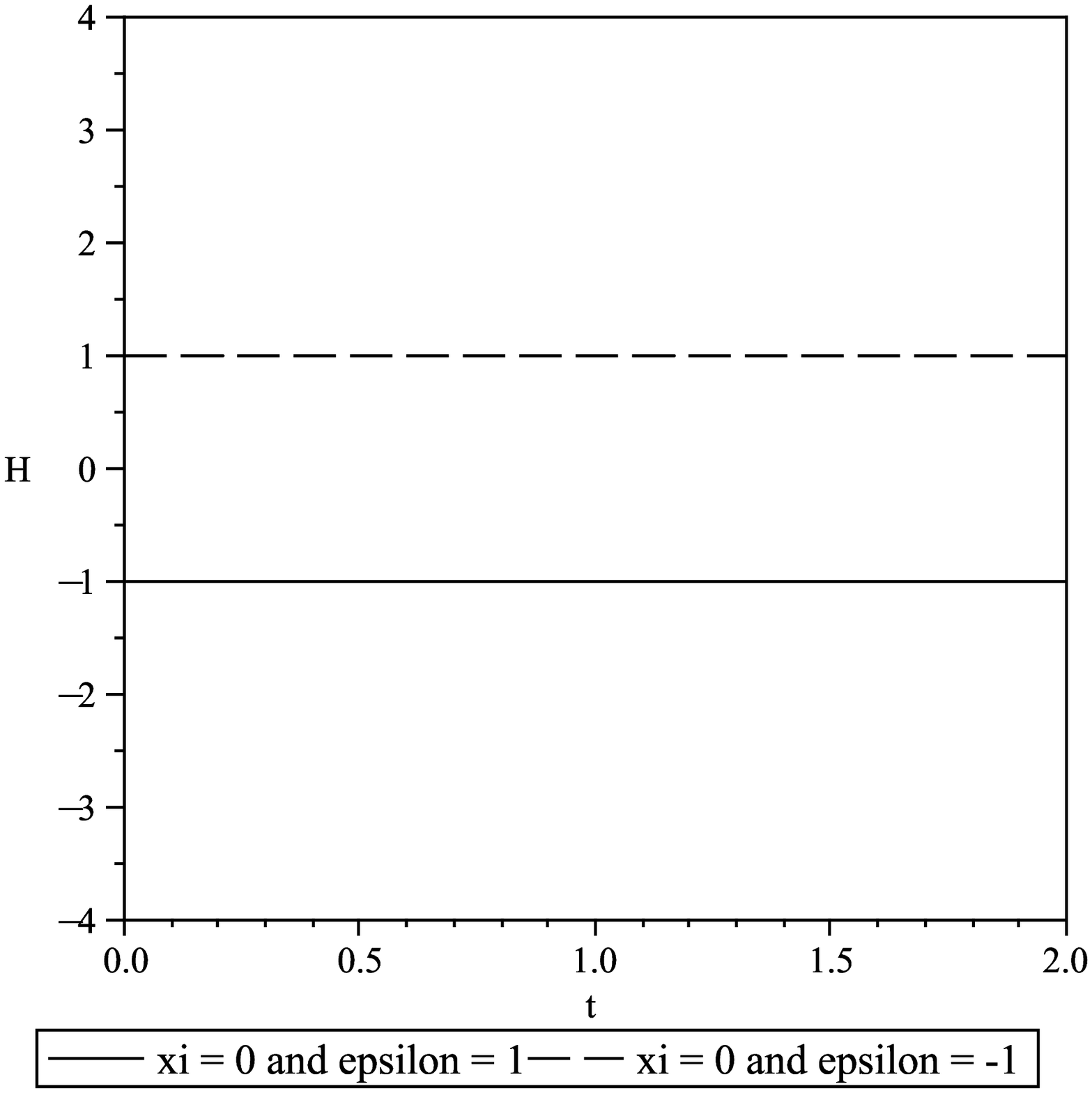} \vspace{5.5cm}\includegraphics{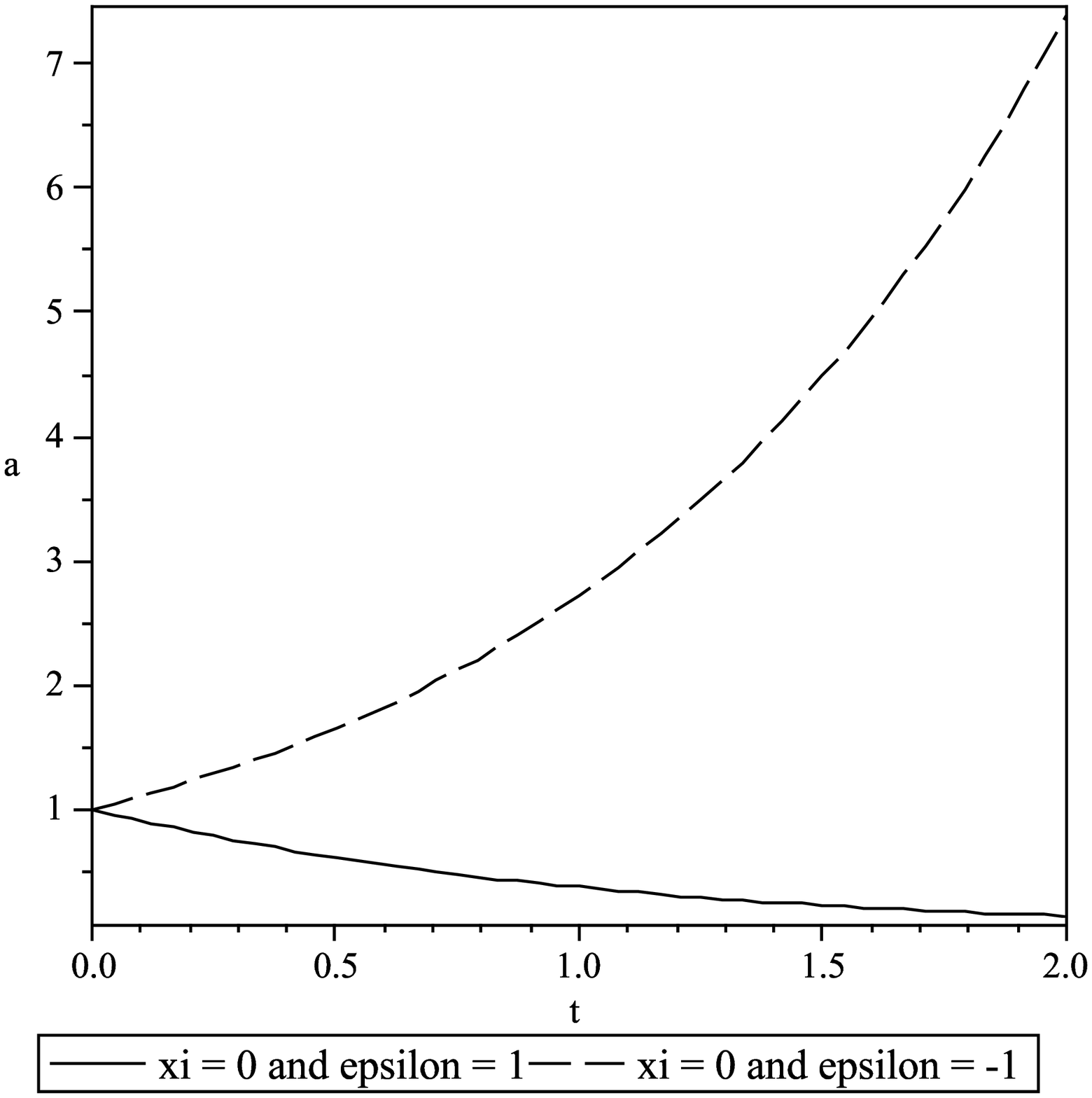}
\end{center}
\caption{\small {Variation of the Hubble parameter $H$ and scale
factor relative to the cosmic time $t$. There is no possibility to
realize bouncing solution in this frame.}}
\end{figure}

\section{Summary}
According to existing literature on scalar field dark energy models,
a minimally coupled scalar field is not a good candidate for dark
energy since its equation of state parameter has not the capability
to realize crossing of the phantom divide line. On the other hand, a
scalar field non-minimally coupled to gravity in the Jordan frame
has the capability to be a suitable candidate for dark energy which
provides crossing of the phantom divide line and other required
facilities. Here we have shown that while a non-minimally coupled
scalar field in the Jordan frame is a suitable dark energy component
with capability to realize phantom divide line crossing, its
conformal transformation in the Einstein frame has not this
capability. In fact, conformal transformation from Jordan frame to
the Einstein frame transforms the equation of state parameter of
dark energy component to its minimal form with a redefined scalar
field and in this case it is impossible to realize a phantom phase
with possible crossing of the phantom divide line. On the other
hand, one of the most serious shortcomings of the standard cosmology
is the problem of initial ( and possibly final) singularity. In
recent analysis done within the loop quantum cosmology, the Big Bang
singularity is replaced by a quintessence Big Bounce with finite
energy density of the scalar field. As we have shown here, in the
Jordan frame with a non-minimally coupled quintessence field, one
achieve a phenomenologically viable framework to overcome initial
singularity with possible realization of the bouncing solutions. We
have studied the stability of this bouncing solutions too. As a
result, there are appropriate regions of the $\omega-\omega'$ phase
plane that solutions are stable in the Jordan frame. However, the
redefined scalar field in the Einstein frame cannot realize these
bouncing solutions.\\

{\bf Acknowledgment}\\
This work has been supported partially by Research Institute for
Astronomy and Astrophysics of Maragha, Iran.

\end{document}